# Opportunities and Challenges for Next Generation Computing

Gregory D. Hager, Mark D. Hill, and Katherine Yelick

Oct. 19, 2015

Version 1

Computing has dramatically changed nearly every aspect of our lives, from business and agriculture to communication and entertainment. As a nation, we rely on computing in the design of systems for energy, transportation and defense; and computing fuels scientific discoveries that will improve our fundamental understanding of the world and help develop solutions to major challenges in health and the environment. Computing innovations "at the high end" tend to "trickle down," leading to increased performance and new applications of computing throughout the entire performance spectrum. These advances have relied on computing innovations in the broadest sense: faster algorithms, new mathematical and statistical models, powerful programming abstractions, ubiquitous high performance networks, and computing systems that have become smaller, faster, cheaper and more accessible over time.

Computing has changed our world, in part, because our innovations can run on computers whose performance and cost-performance has improved a million-fold over the last few decades. A driving force behind this has been a repeated doubling of the transistors per chip, dubbed Moore's Law. A concomitant enabler has been Dennard Scaling that has permitted these performance doublings at roughly constant power, but, as we will see, both trends face challenges.

Consider for a moment the impact of these two trends over the past 30 years. A 1980's supercomputer (e.g. a Cray 2) was rated at nearly 2 Gflops and consumed nearly 200 KW of power. At the time, it was used for high performance and national-scale applications ranging from weather forecasting to nuclear weapons research. A computer of similar performance now fits in our pocket and consumes less than 10 watts. What would be the implications of a similar computing/power reduction over the next 30 years – that is, taking a petaflop-scale machine (e.g. the Cray XK7 which requires about 500 KW for 1 Pflop (=$10^{15}$ operations/sec) performance) and repeating that process? What is possible with such a computer in your pocket[1]? How would it change the landscape of high capacity computing? How would it change the landscape of personalized computing? Will such computing be general purpose and programmable in the same way we're accustomed, or will new paradigms need to emerge?

While such performance improvements do not guarantee the same revolutionary changes we've seen over the past 30 years, they dramatically change the landscape of possibilities for both national-scale and personalized computing. In the remainder of this paper, we articulate some opportunities and challenges for dramatic performance improvements of both personal to national scale computing, and discuss some "out of the box" possibilities for achieving computing at this scale.

---

[1] One might also ask how such a computer compares to the one in your head! As a point of comparison, the human brain is estimated to contain around $8 \times 10^{10}$ neurons and $10^{14}$ synapses. Neural circuits operate at firing rates of 100 to 1000 Hz. Although direct comparisons are hard to make, this could be viewed as 10-100 "petaspikes" of computation operating on around 20 watts of power.

## Large Scale Computing Challenges

The list below offers suggestions of problems that will be shaped by future advances from computing at this scale. This list spans problems from traditional modeling and simulation, to advanced data analysis and the interface between simulation and observation, looking for both advances in traditional computing systems and in opportunities for discoveries that use new models of computing.

**Anticipating extreme weather events through modeling and monitoring**

Modern climate models predicate long-term global and regional changes in temperature and precipitation, but do not adequately simulate extreme regional events such as hurricanes or droughts. In contrast, weather prediction models use observational data at regional scales to forecast weather up to a week in advance. Specific storm paths or extreme temperature are predicted in a probabilistic sense. Hundreds of times more computing is required to reach the resolution at which cloud systems can be explicitly resolved and precipitation more accurately modeled at a regional scale. In addition, real time data from satellites, carbon sensors, and weather monitoring needs to be de-noised, analyzed and integrated to improve the models and prediction capabilities. The climate simulations will use current digital computing models, although more specialized architectures or new low power device technology will be needed to move to credible projections of future changes in extreme weather. In contrast, image and sensor analysis is not done automatically at scale today, and may take advantage of new computational models that are better at identifying patterns and errors both within and across different types of data. Climate analysis is only one of many examples within science and engineering in which the observational data and simulation will be combined to improve predictive capabilities.

**Understanding quantum effects in materials and chemistry models**

Advances in models, algorithms and classical high performance computing systems have made simulation an invaluable tool in chemistry and material science, with scientific and commercial applications. Massive high throughput computing in the Materials Genome Initiative allows for rapid screening and the identification of materials with particular properties. But understanding detailed characteristic of chemical and material systems, such as low-lying electronic states, chemical reaction dynamics, or electron transfer processes at experimental precision is intractable for large molecular or condensed matter systems. These are important in problems such as nitrogen fixing and understanding photosynthesis. Quantum simulations exhibit exponential scaling in computational complexity when approaching accurate solutions of the quantum mechanical equations, but simulation on quantum devices has shown promising results for small molecules on small numbers of qubits, and modest sized quantum devices (tens of qubits) could be used to simulate larger systems that are intractable on classical computer systems. The process of how to control and program a quantum device at this scale is not well understood, nor is there a solid theoretical foundation in computing to understand whether there is an inherent complexity advantage in quantum devices for simulation problems.

**Search engines for science**

New data policies are making previously private scientific data available to the community, enabling new modes of scientific discovery that involve combining data across disciplines, reanalyzing data collected for another purpose, and improving confidence in through larger analyzes. Rather than finding data manually by, for example, knowing a particular paper or author, imagine a search engine that discovers metadata information, extracts features from images and other scientific data sets, summarizes and combines relevant articles through natural language processing, and infers models that can be used for simulation and prediction. For example, one could deduce that a compound or new drug has toxicity without having to perform experiments on mice, or identify interesting transients in a stream of images to automatically control telescopes and microscopes. Search engines that we take for granted in shopping, travel planning, or reading the news could be available for scientific data. Rather than exploring each hypothesis one at a time, tools that infer relationships and identify structure should allow scientists to quickly explore a space of interpretations.

**Prediction of human-in-the-loop systems**

Some of the most interesting and important systems in societal challenges involve understanding and predicting human behavior. For example, how will people react to changes in energy pricing, new ingredient or labeling on foods, or news about a financial or health emergency? While aggregate statistical models exist for global behavior, e.g., the spread of disease, they are not detailed enough to manage energy infrastructure, control traffic or make specific medical decisions. Learning these models from past experience would enable better prediction and optimization of systems such as smart grids for power distribution, or smart cities with integrated energy production and saving devices.

## Individualized Computing Paradigms

The computing advances of today – combining powerful local processing with enormous "backend" cloud capabilities has enabled a revolution in sense-making from textual data and sensory data processing – speech, vision, and sound analysis. Indeed, most of today's most compelling applications – speech recognition, face detection and recognition – rely on the fact that the cloud-based systems are able to train algorithms on enormous data sets, and run those algorithms at scale once trained.

*Locally Adaptable Systems:* Today's computing devices are highly flexible because they are programmable. But, programming typically presumes customization of function at scale – software is designed to run on millions of cars, phones, laptops, or refrigerators. Conversely, machine learning from data can potentially capture individual trends with high fidelity. But, the current process of training from data is compute intensive, and requires a human to frame the problem and to tune the algorithm to achieve high levels of performance. What if data modeling and machine learning was available in every future computing device?

The next generation of performance gains could make "learning from data" capabilities available in a power-efficient microprocessor-sized package format. How might this change the way systems would be developed? The microprocessor in your car could now monitor your driving behavior, recording video and other sensory data over days and weeks, and then train itself to simulate the type of driving you are comfortable with, and react appropriately to the behavior of drivers on the road in your area. Your home surveillance system could observe the activities in your home and learn a model of how you and your family behave to better detect and warn about anomalous behavior. Your phone could learn to anticipate when you are going to lunch, your standing meetings, and the people you interact with, and help you manage or structure your day. Indeed, every device could contain multiple "learning" systems that would adapt to their surroundings and use in ways designated by the end-user.

*Local Modeling:* There are many functions (e.g. weather forecasting) that currently take place offline and at scale because they are too complex to perform on demand and for a local situation. But, petaflop-scale computing could, in principle, be used to, for example, aggregate data from your local environment to produce a fine-scale minute-by-minute weather forecast for your morning run, or a drive to a neighboring city, or the beach near the hotel where you are staying. It could monitor your health data, compare you to the population in your area, and warn you about potential health events, or guide you toward healthier behaviors.

Likewise, buildings, bridges, roads, and most other infrastructure are heavily simulated before they are built to ensure safety and reliability. Once built, simulations are no longer used. What if high fidelity simulations were built into major infrastructure, and the simulations compared against sensor data from the infrastructure over time? Would it be possible to detect or predict failures long before they happen based on deviations from the model?

Taken together, these capabilities could also comprise a next generation of "front-end" data structuring for future supercomputers performing climate change modeling, national infrastructure modeling and costs predictions, and many other large-scale, national or global predictive models.

*Real-time Simulation:* Simulation is rapidly becoming a key technology for advances in robotics, cyberphysical systems, and other areas where structured interaction with the surrounding world is necessary. Currently, as with weather forecasting, nuclear simulation, or other physical processes, the fidelity of those simulations is limited by computing. Petaflop scale simulation would be transformational to these systems as high-fidelity, real-time simulation could be integrated directly into the control and decision loops of these systems. These simulations could be directly coupled to sensing processes, facilitating high-reliability and highly-general interaction with the physical world.

## Enabling Computing Outside the Box

Over the last decade, however, Moore's Law has slowed, while Dennard Scaling has ended. Consequently, computer systems can keep doubling the number of active transistors only when one can repeatedly double the power used, which is not viable anywhere: not in the handheld, datacenter, or supercomputer. Thus, new approaches are needed to enable the next generation of computing innovations. These might range from clever use of techniques which are already emerging today to entirely different paradigms both for producing physical devices themselves and for programming their function once produced. Indeed, computing may eventually include physical phenomena that mirror processes being simulated, modeled, or controlled. We describe a few ideas below.

**Next Generation Heterogeneous Computing**

Computer systems can contain more transistors, provided that only carefully chosen subsets are active at any one time, while the rest are *dark silicon*. A key challenge for computer design is how to best use dark silicon to continue doubling performance or cost-performance. A leading candidate for this is *heterogeneous computing* with conventional cores in concert with *accelerators*. Accelerators are specialized hardware that performs a subset of computing tasks faster and/or more energy-efficiently than conventional general-purpose computing. This idea is not new—floating-point arithmetic was part of an early accelerator, and graphics processing units (GPUs) are a recent example—but a broader range of accelerators make even more sense today for two reasons that follow from today's focus on energy. First, accelerators often improve energy-efficiency even more than performance. Second the energy opportunity cost of accelerators not in use can be made near zero by turning such accelerator off. As never before, it is now viable to have a palette of accelerators with just a few in use at any one time.

The design of accelerators is challenging. What function should accelerators target? Beyond graphics, recent accelerators have examined encryption/decryption, compression/decompression, database processing, artificial neural networks, and facial recognition. More generally, how specific should accelerators be? Should they do relatively specific tasks like the list above, very flexible like programmable fabrics like field-programmable gate array (FPGAs), or should they be some configurable option in the middle? There are important tradeoffs here that will require communication among problem domain experts, hardware designers, and software experts.

The logical and physical integration of accelerators with the rest of the computer system is another critical tradeoff. At one extreme, accelerators can be joined with general-purpose cores like floating-point or some vector/single-instruction-multiple-data (SIMD) hardware. This option lowers the cost of transfer work to and from an accelerator but such close integration is a precious resource to be husbanded. At the other extreme, accelerators can be placed on the input/output (I/O) fabric like conventional discrete GPUs. Here many low-use accelerators may be included but these accelerators must operate on relatively coarse-grain tasks to amortize the effort of transferring data to and fro. An emerging, attractive comprise is attaching accelerators to the memory interconnect like some direct-memory-access (DMA) engines. This middle option can allow many accelerators access to all of host memory with the potential for high performance and more general programmability (e.g., using the same data structure as programs running on host cores without copying or reworking pointers).

The programming model and software tool chain for accelerators is also critical. As we have learned before for vectors and parallelism, broad success of new hardware requires appropriate programming abstractions and tools, as well as a substantial performance benefit. One open question is whether accelerators will be automatically invoked by compilers and runtime systems, hidden within domain-specific libraries written by performance experts, or be explicitly managed by application programmers. The details of the architecture will be important, such as how easily data can be shared across accelerators and CPUs, the amount of state within the accelerator, and the availability of tools to manage and optimize their use.  The breadth of algorithms that can take advantage of the architecture is also an essential factor in building and sustaining a robust ecosystem of programmers; this was apparent in the use of GPUs for scientific computing, which leveraged languages (like CUDA) and a programming community (including many students) that were outside of traditional scientific computing. A challenge with accelerators will be the desire for portability both in functionality and in performance, i.e., the ability to write a single application that will run well on many different systems, and whether this goal can be met via general-purpose languages (liked C++ or Python), accelerator-specific languages (like CUDA), domain-specific languages (e.g., for linear algebra, stencils, or spectral algorithms) or hidden in parameterized libraries.

**Beyond Today's Box**

While heterogeneous computing appears the likely successor to today's computing, it is also critical that we lay the foundation for more radical *out-of-the-box* alternatives. We consider below four aspects of the current "box" and how to break out.

One aspect of the current box is that most computer systems separate processing, memory, communication, and storage. This separation partitions complexity (divide and conquer) and facilitates use of different technologies (from CMOS to magnetic disk drives). This separation, however, comes with the energy cost of moving information among components. Fortunately, new technologies may enable breaking out of this box. For example 3D die stacking allows computational and memory processes to be physically and logically close, while emerging byte-addressable non-volatile memory technologies can blur the distinction between memory and storage. Moreover, many emerging "big data" problems may facility new models where a small quantum of data (e.g., a query or item to match) can move to interact with a large, stationary corpus.

Another aspect of the box follows from near-exclusive use of high-precision digital values, such as 64-bit double-precision floating point. This use makes sense for performance, because today's cores add and multiply such numbers with ease. However, it takes energy to do these computations and, more importantly, to move and store the numbers. On the other hand, data in many emerging applications can be noisy on input (from sensors) or only low-precision output is necessary (some graphics). One can break out of this box with better support for low-precision or even analog values. More radically, direct support for probabilistic values and computation may save conventional processing steps on some problems, such as statistical inference or in machine learning.

A foundational box of computing is that of logical time steps. Computing today makes ubiquitous use of abstractions like finite-state, Turing, random-access machines. All of these have the notion that the current state and inputs determine the next state and outputs. Time is abstracted away and implementations seek to go as fast as possible using a synchronous high-frequency (gigahertz) clock or delay-variation-tolerant asynchronous handshake protocols. To break out of this box, we should consider designs that seek not to hide time but rather to use it to represent information. The key argument for considering this is that spiking times in the human neocortex seem well coordinated and likely to represent part of the brain's encoding of information. This aspect should be investigated, given the brain's capabilities and energy efficiency.

A final aspect of today's box is that most computers today rely on classical interactions. In contrast, quantum computers offer efficiency for aspects of some important computations (e.g., the materials/chemistry simulations mentioned earlier). A key question might be: will quantum computers remain stand-alone or can they be integrated into classical computers as accelerators? The advantage of integration is that only some aspects of computation benefit from quantum acceleration and doing just these may require fewer quantum bits (qubits) than stand-alone quantum computers. The very big disadvantage is that such integration requires effective operation at more normal temperatures than current quantum computers.

## Summary

The future is unknowable. History teaches us that applications of computing will be shaped by as yet unanticipated changes and broad societal trends. Indeed, many of current impacts of computing were hard to anticipate more than a decade prior – while mobile devices, embedded computers, the world-wide web, and high performance computing centers were evident, the widespread adoption and transformative nature of these systems was not apparent.

But, we do know that there is an immense reservoir of innovation possible if computing performance continues to advance at all performance levels. The challenges outlined above frame some of these opportunities. To achieve them, we should invest in a diverse portfolio to further enable more performance and cost-performance growth. Aspects of heterogeneous computer represent the "blue chip" component of this portfolio, while out-the-box ideas are like startup companies where most will fail but some will succeed and by doing so cause major disruptions that will fuel the next generation of computing-based innovations for our society.

*For citation use*: Hager G. D., Hill M. D., & Yelick K. (2015). *Opportunities and Challenges for Next Generation Computing*: A white paper prepared for the Computing Community Consortium committee of the Computing Research Association. http://cra.org/ccc/resources/ccc-led-whitepapers/

This material is based upon work supported by the National Science Foundation under Grant No. (1136993). Any opinions, findings, and conclusions or recommendations expressed in this material are those of the author(s) and do not necessarily reflect the views of the National Science Foundation.